\documentclass{iopjournal}

\usepackage[normalem]{ulem}

\usepackage{amsmath}
\usepackage{amssymb}
\usepackage{hyperref}
\usepackage[T1]{fontenc}
\usepackage{xcolor}
\usepackage{upgreek}
\usepackage{cuted}
\usepackage{graphicx}
\usepackage{bm}
\usepackage{cite}

\begin{document}

\articletype{Paper} 

\title{Confinement-Induced Pumping of Chiral Active Fluids}

\author{Joscha Mecke$^{1,2}$\orcid{0000-0001-8984-6990}, Yongxiang Gao$^2$\orcid{0000-0003-4042-0248} and Marisol Ripoll$^{3}$\orcid{0000-0001-9583-067X}}

\affil{$^1$Institut f\"ur Theoretische Physik II - Weiche Materie, Heinrich-Heine-Universit\"at D\"usseldorf, Universit\"atsstraße 1, D-40225 D\"usseldorf, Germany}

\affil{$^2$Institute for Advanced Study, Shenzhen University, 518060 Shenzhen, P.R.~China}

\affil{$^3$Theoretical Physics of Living Matter, Institute for Advanced Simulation, Forschungszentrum J\"ulich, 52425 J\"ulich, Germany}

\email{joscha.mecke@hhu.de}
\email{yongxiang.gao@szu.edu.cn}
\email{m.ripoll@fz-juelich.de}


\begin{abstract}
Chiral active fluids, composed of continuously rotating active constituents, exhibit a wealth of nonequilibrium phenomena arising from the interplay of activity, hydrodynamic interactions, and broken mirror symmetry. While these systems naturally generate circulating flows and active turbulence, converting their microscopic rotational motion into directed macroscopic transport remains an open challenge. By means of mesoscale hydrodynamic simulations, here we show that channels with an intrinsic asymmetry result in the spontaneous formation of a persistent net flux in the absence of externally imposed pressure gradients or body forces. 
Our results can be extrapolated to various geometric confinement strategies providing efficient and versatile strategies for transforming chiral activity into autonomous fluid transport, such that they can be utilized as a promising platform for self-powered microfluidic pumping.
\end{abstract}

\section{Introduction}
Fluids consisting out of building blocks capable of consuming energy at the microscopic scale in order to apply forces or torques on the surroundings are known as active fluids~\cite{ramaswamy2019active, marchetti2013hydrodynamics} with examples in biological systems, such as bacterial suspensions~\cite{aranson2022bacterial}, as well as synthetic active matter~\cite{theuerkauff2012dynamic}. %
Depending on the mode of activity and the interactions between the active agents a wealth of emergent phenomena, including flocking~\cite{yang2015hydrodynamics}, active turbulence~\cite{wensink2012meso}, or directed motion~\cite{bricard2013emergence} with no counterpart in passive fluids have been reported~\cite{gompper2025roadmap}. %

A particularly intriguing class of active matter is formed by chiral active fluids, whose constituents continuously inject angular momentum into the surrounding medium through persistent rotation~\cite{liebchen2022cam, mecke2024emergent}. Such systems can be realised experimentally using externally driven magnetic~\cite{katuri2024control, gardi2022microrobot} or optical colloidal rotors~\cite{modin2023hydrodynamic}, as well as by intrinsically chiral microorganisms~\cite{riedel2005self} and synthetic microswimmers~\cite{mousavi2020bacteria}. 
The rotational motion of the individual particles generates circulating solvent flows that hydrodynamically couple neighbouring rotors transversally over long distances~\cite{mecke2023simultaneous, modin2023hydrodynamic}. Depending on the particle concentration and activity, these interactions lead to a broad spectrum of collective states, ranging from dilute Brownian suspensions to active turbulence characterised by multiscale vortical structures~\cite{reeves2021emergence}. In contrast to conventional active fluids, where force dipoles dominate the dynamics, chiral active fluids are governed by active torques and antisymmetric stresses, resulting in transport phenomena that are unique to systems with broken mirror symmetry~\cite{chatzittofi2025dumbbell, mecke2024emergent, fruchart2023odd, hargus2025odd, kalz2022collisions} and differ fundamentally from those of translationally driven active matter~\cite{hosaka2023lorentz, kalz2026reversal, soni2019odd}. %

Under conditions of confinement, active fluids show the emergence of spontaneous symmetry breaking~\cite{zhang2022polar, nishiguchi2025vortex}, pinning of emergent flow structures~\cite{reinken2020organising}, or confinement induced pumping~\cite{varghese2020confinement}. %
Particularly, in confined chiral active fluids, recent theoretical and experimental studies have demonstrated the emergence of robust edge currents~\cite{dasbiswas2018topological, li2024robust, mecke2025obstacle, abdoli2026ddft, caprini2026active} featuring unidirectionality set by the system's chirality. %
Accordingly, it is possible to tailor systems with preferred directions of motion~\cite{ai2015chirality} and stabilising collective dynamical structures~\cite{caprini2025spontaneous}. %
These observations suggest that geometrical confinement may provide a powerful means of controlling emergent flows~\cite{reichhardt2017ratchet, zhang2022polar} generated by chiral activity. %

A fundamental question, however, remains open: can confinement alone convert the intrinsically circulating motion of a chiral active fluid into a directed net flow? In an unbounded system, the rotational symmetry of the active stresses precludes any preferred transport direction, and the time-averaged flux vanishes. Channel geometries break translational and rotational symmetries through the presence of confining walls and therefore offer a minimal setting in which microscopic chiral motion may be rectified into macroscopic pumping. Understanding this mechanism is relevant not only for the fundamental hydrodynamics of chiral fluids but also for the design of self-powered microfluidic pumps that operate without externally imposed pressure gradients or moving mechanical components.

In this work, we investigate pressure-free pumping generated by a confined chiral active fluid in channel geometries. %
We study a quasi-2D system of colloids rotating at constant angular velocity following an explicit experimental setup~\cite{mecke2023simultaneous, mecke2024substrate} of a monolayer of hydrodynamically coupled magnetic particles exposed to an externally applied rotating magnetic field. %
We employ explicit-solvent multiparticle collision dynamics simulations of hydrodynamically interacting colloidal rotors, using the same experimentally validated model that has previously been shown to reproduce the collective dynamics of dense chiral active suspensions. %
By systematically varying the channel geometry and the colloidal density of the suspension, we identify the physical mechanism responsible for the emergence of a net flux. We demonstrate how the interplay between active edge currents, hydrodynamic interactions, and channel confinement produces spontaneous pumping, and we characterise the dependence of the resulting flow on the system parameters. Our results establish asymmetric channel confinement as a simple and robust strategy for converting microscopic chiral activity into directed macroscopic transport.

\section{Methods}
We simulate the colloidal chiral active fluid by means of an ensemble of rotating colloids suspended to a two-dimensional explicit solvent where hydrodynamic interactions are fully resolved. %
The model follows our previous work closely mimicking experiments of a carpet of standing and synchronously spinning colloidal rods~\cite{mecke2023simultaneous, mecke2024substrate, mecke2025obstacle}. 
The solvent is accounted by multiparticle collision dynamics (MPC)~\cite{kapral2008multiparticle,ripoll2005dynamic,gompper2009multi} and consists of point-like particles of mass $m$ and density $\rho$ whose positions $\bm{r}_i$ and velocities $\bm{v}_i$ are updated at discrete time steps of time step-size $h$.
In each time step, each fluid particle first streams ballistically $\bm{r}_i(t+h) = \bm{r}_i(t) + \bm{v}_i(t)h$. %
Subsequently, the particles are sorted into square collision cells of box length $a$ and interchange momentum with the $N_\zeta$ particles in the same cell according to a rule ensuring the conservation of mass, linear and angular momentum via
\begin{align}
	\bm{v}_i(t+h) = \bm{v}_\zeta(t) + \begin{pmatrix} \cos\vartheta & -\sin\vartheta \\ \sin\vartheta & \cos\vartheta \end{pmatrix} \cdot (\bm{v}_i(t) - \bm{v}_\zeta(t)) - (\bm{I}_\zeta^{-1}\cdot\Delta\bm{L}_\zeta) \times (\bm{r}_i(t) - \bm{r}_\zeta)\,,
\end{align}
where $\bm{r}_\zeta = \sum_{i\in\zeta} \bm{r}_i/N_\zeta$ denotes the centre-of-mass position in a given collision cell $\zeta$ and similarly $\bm{v}_\zeta$, $\bm{I}_\zeta$, and $\Delta\bm{L}_\zeta$ denote the centre-of-mass velocity, moment of inertia, and change in angular momentum in a given cell, respectively~\cite{noguchi2008transport}. %
The correction term containing $\Delta\bm{L}_\zeta$ ensures angular momentum conservation such that no unphysical torques between the rotating colloids emerge~\cite{gotze2007relevance, pooley2005kinetic}. %
In order to avoid strong correlations in the velocities of the fluid particles in the same collision cell over consecutive time steps, a random shift of the collision cell grid is performed before each collision~\cite{ihle2001stochastic}. %

In experimental realisations of quasi-two-dimensional sheets of chiral active fluids, influences of the third dimension can lead to an attenuation of the created flows caused by finite friction between the active fluid layer and nearby substrates or interfaces~\cite{soni2019odd} which can be successfully modelled by a linear damping term proportional to the friction coefficient $\gamma$. %
The friction scale can vary from a few to tens or hundreds of rotor diameters and can lead to a localisation of emergent flows~\cite{dasbiswas2018topological}. %
The interaction of the fluid with an underlying frictious substrate is modelled by employing an additional set of virtual particles scattered over the simulation domain with velocities drawn from a Maxwell-Boltzmann distribution. The virtual particles take part in the collision routine, thus interchange momentum with the fluid particles, and hence act as a momentum sink. In this way, we can accurately incorporate the (damping) influences of the third dimension on the quasi-2D fluid sheet with a friction coefficient~\cite{mecke2024substrate}
\begin{align}
	\gamma = \frac{1}{h} \frac{\rho_\mathrm{s}}{\rho + \rho_\mathrm{s}} \left( 1 - \frac{1}{2(\rho-1)} \right)
\end{align}
where $\rho$ and $\rho_s$ are respectively the average density of fluid and virtual particles per collision box, with $\rho_s \ll \rho$.  
Here we have assumed that $\vartheta=\pm\pi/2$, and corresponding with this friction coefficient the typical friction decay length is $\lambda = \sqrt{\nu/\gamma}$, with $\nu$ the fluid's kinematic viscosity. %
This approach has been successfully applied in active two-dimensional fluids in order to regularise 2D hydrodynamics at large distances, resolving Stokes' or Jeffery's paradoxes~\cite{mecke2023birotor}. %

The rotating colloids are modelled as circular no-slip boundaries exchanging angular and linear momentum with the MPC particles. During streaming, the momentum transfer takes place according to a bounce-back protocol~\cite{gotze2011dynamic}. During the collision step, the fluid particles interchange momentum with a set of virtual particles sampled on the inside of the colloids, which are sorted by position into the same collision cells as the fluid particles. %
After the collision, the momentum change of the virtual particles is assigned to the colloids centre-of-mass linear and angular velocities~\cite{gotze2011dynamic}. %
The colloids rotate with a fixed angular velocity modelling phase locked dipolar colloids in a rotating field~\cite{mecke2023simultaneous, soni2019odd}. %
We employ a cell-level thermostat~\cite{huang2010cell} in order to prevent energy accumulation resulting from the constant activity input, guaranteeing a constant system temperature $k_\mathrm{B}T$. %
The steric interactions between colloids are modelled with a shifted WCA-potential
\begin{align}
	U(r) = \begin{cases} 4\epsilon \left[ \left( \frac{a}{r-\sigma} \right)^{12} - \left( \frac{a}{r-\sigma} \right)^6 \right] +\epsilon\,, & \text{for } r\le\sigma+2^{1/6}a\\ 0\,, & \text{else.} \end{cases}
	\label{eq:wca}
\end{align}
This potential also ensures that  there is always at least one collision cell between approaching colloids and a colloid and the wall ensuring proper hydrodynamic coupling. %
The walls are modelled as non-movable no-slip boundaries and the colloids are similarly repelled from the walls according to a shifted WCA-potential when approaching closer than $r\le\sigma/2+2^{1/6}a$. %
The positions and velocities of the colloids are updated using a molecular-dynamics Verlet scheme~\cite{allen2017computer} with a time increment $h$. %

For the MPC-fluid, we employ a density of $\rho = 10\, m/a^2$, a rotation angle of $\vartheta = \pm\pi/2$ with equal probability, a collision time $h=0.02\, a\sqrt{m/k_\mathrm{B}T}$ and thus obtain a viscosity of $\eta = 17.9\sqrt{mk_\mathrm{B}T}/a$. %
We fix the colloidal diameter to $\sigma=6a$ and rotational velocity to $\Omega = 0.01/(a\sqrt{m/k_\mathrm{B}T})$ leading to a rotational Reynolds and P\'eclet numbers of $\mathrm{Re} = \sigma^2\Omega/\nu = 0.1$ and $\mathrm{Pe} = \sigma^2 \Omega/D = 27$, respectively, where $D$ is the colloid's diffusion coefficient. %
The virtual substrate particle density, unless otherwise stated, is $\rho_\mathrm{s} = 0.00005 m/a^2$ resulting in $\lambda = 14.5\sigma$. %
In the experiments in Refs.~\cite{mecke2023simultaneous, mecke2024substrate} closely related to the system considered here, the friction scale can be estimated as $\lambda \simeq 11.3\sigma$. %
We employ standard MPC units as $m=a=k_\mathrm{B}T=1$. %
The dynamics of the up to several thousand colloids and millions of fluid particles is simulated employing a custom massively parallelised GPU code running on the JUWELS supercomputer~\cite{kesselheim2021juwels}.

\section{Results}

\subsection{Net flow generation}
In bulk conditions, each spinning colloid (rotor) creates a co-rotating flow in the surrounding solvent decaying like $u \propto r^{-1}$, which results from the no-slip boundary condition on the surface of the colloids. %
The slow decay holds at distances shorter than the typical friction decay length, $r\ll\lambda$, at large distances a Bessel cutoff resulting from the substrate friction $u \propto K_1(r/\lambda)$   
dominates, where $K_1(r)$ is the modified Bessel function of the second kind. %
While isolated rotors do not show any translational activity and merely perform a random walk, two approaching rotors will be dragged by each other engaging in an orbital rotation about the centre-of-mass. The internal rotation is thus coupled to the translational degrees of freedom through the hydrodynamic coupling, leading to collective motion featuring the formation of vortices and eddies of various sizes, a behaviour that has been characterised as (chiral) active turbulence~\cite{mecke2023simultaneous}. %
On a coarse-grained level, the rotation-translation coupling can be described as a rotational stress proportional to the rotational viscosity $\eta_\mathrm{R}$, \textit{i.e.}, $\sigma^\mathrm{rot}_{\alpha\beta} = \varepsilon_{\alpha\beta}\eta_\mathrm{R}(2\widetilde{\Omega} - \omega)$, where $\widetilde{\Omega}\sim \Omega \phi$ is the angular velocity density, $\phi$ the colloidal area fraction, $\omega = \varepsilon_{\alpha\beta}\partial_\alpha v_\beta$ is the vorticity of the coarse-grained colloidal flow~\cite{mecke2024emergent} and $\varepsilon_{\alpha\beta}$ is the Levi-Civita symbol. %
$\sigma^\mathrm{rot}_{\alpha\beta}$ acts as a synchronisation between internal rotation and rotational flow, \textit{i.e.}, vorticity. %

\begin{figure}[h]
	\centering
	\includegraphics[width=\textwidth]{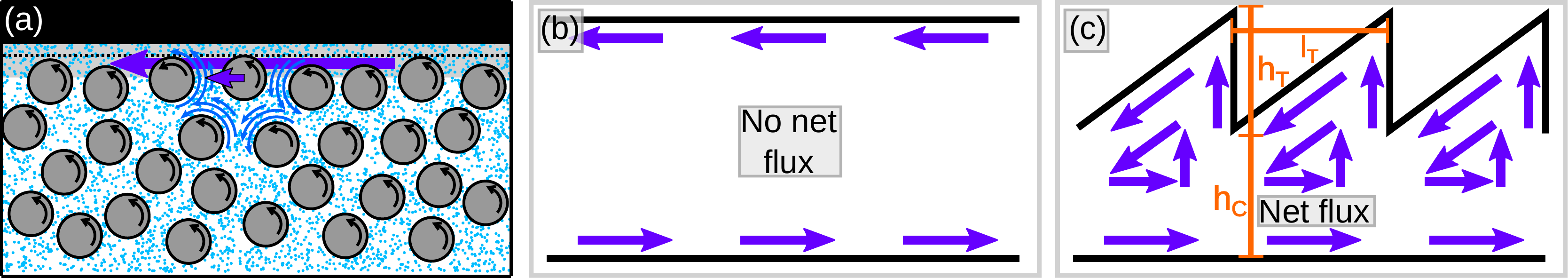}
	\caption{Chiral edge modes facilitate net transport in asymmetric channels.
		(a)~Schematic of the chiral fluid at a wall. Colloidal rotors are depicted in gray, solvent particles in blue, and steric wall-colloid interactions by a shaded area representing the excluded volume for the colloids, filled with fluid particles. 
		Rotors circulate in the direction opposite to their inherent direction of rotation essentially since they are being carried by each other’s flows,  as illustrated by the arrows. %
		(b)~Sketch of a channel with planar walls where identical flows are generated in opposite directions  resulting in no net flux. 
		(c)~Sketch of a channel with one ratcheted and one planar wall where the generated flows in opposite directions are different, resulting in a persistent net flux. Relevant channel sizes are indicated. %
	}
	\label{fig:fig1}
\end{figure}
When the active fluid is near a wall, the rotor layer that is closest to the wall will experience unidirectional rotational stresses, leading to the formation of an edge current into the direction opposite to their inherent direction of rotation, as sketched in Fig.~\ref{fig:fig1}a. %
The strength of this edge current can be predicted using a generalised Stokes description featuring $\sigma^\mathrm{rot}_{\alpha\beta}$. The full flow is then prescribed by the generalised Stokes equation and the boundary conditions. %
At the walls, the active fluid experiences stresses resulting from the thin fluid layer between the no-slip boundary conditions on both the surfaces of the colloids and the walls. While the colloids do not directly experience stresses at the wall, the stresses are transmitted via the viscous fluid coupling~\cite{mecke2025obstacle}. %
Generally, the emergent flow will depend on the location along the wall and may vary together with the local geometry. However, we can extend the result from Ref.~\cite{mecke2025obstacle} to the limit of an infinite flat wall under conditions with substrate friction and obtain for the velocity of the edge mode
\begin{align}
	v_\mathrm{edge} = \frac{2\eta_\mathrm{R}\widetilde{\Omega}}{\lambda^{-1}(\eta+\eta_\mathrm{R}) + \eta/\delta}\,,\label{eq:v_edge}
\end{align}
where $\delta$ refers to the minimum thickness of the thin fluid layer between the rotor surfaces and the wall~\cite{mecke2025obstacle}, which in our case is determined by Eq.~(\ref{eq:wca}). %
Note, that edge currents in chiral active fluids have been found to be topologically protected~\cite{souslov2019topological, shankar2022topological, dasbiswas2018topological}, leading to a robust current phenomenon~\cite{li2024robust} that reliably navigates unidirectionally and almost scatter-free around corners and obstacles. %
This ability of chiral active systems has already been exploited for cargo transport along a wall, where the rotor fluid's odd viscosity additionally stabilises the cargos dwelling in the edge flow, \textit{i.e.}, at the wall~\cite{yang2021topologically, yang2021edge}. 

When confined between two parallel walls, the edge currents generated at both walls will be of equal strength and opposite directions, as illustrated in Fig.~\ref{fig:fig1}b. 
Consequently, no net flux is obtained for a symmetric linear channel~\cite{gotze2011dynamic}. %
However, if the upper and lower walls are not symmetric, as those in Fig.~\ref{fig:fig1}c, the emergent flows into the $x$-direction created at both walls may not cancel. The flow still exhibits a stagnation point, however, it does not necessarily lie in the middle of the channel and thus admits to net flux into the $x$-direction.

\subsection{Asymmetric jigsaw geometry}
We consider a channel with a jigsaw-shaped wall and a flat wall opposite (see Fig.~\ref{fig:fig1}c). %
We simulate the chiral active fluid in a domain of five identical tooth compartments of length $l_\mathrm{T}=34\sigma$ and apply periodic boundary conditions in the $x$-direction, such that the total channel length is $L_x= 170\sigma$. %
The tooth-height is varied as $h_\mathrm{T}/\sigma = 3, 10, 30, 45, 48$, while the total channel height is kept constant $H=h_\mathrm{T} + h_\mathrm{C}$, such that the ratcheted wall is inclined at a variable angle $\theta=\arctan (l_\mathrm{T}/h_\mathrm{T})$. %
Unless otherwise stated, a rotor density of $\phi=N\pi(\sigma/2)^2/(L_x H -  5l_\mathrm{T} h_\mathrm{T}/2) = 0.2$ was employed. %

\begin{figure}[t]
	\centering
	\includegraphics[width=\textwidth]{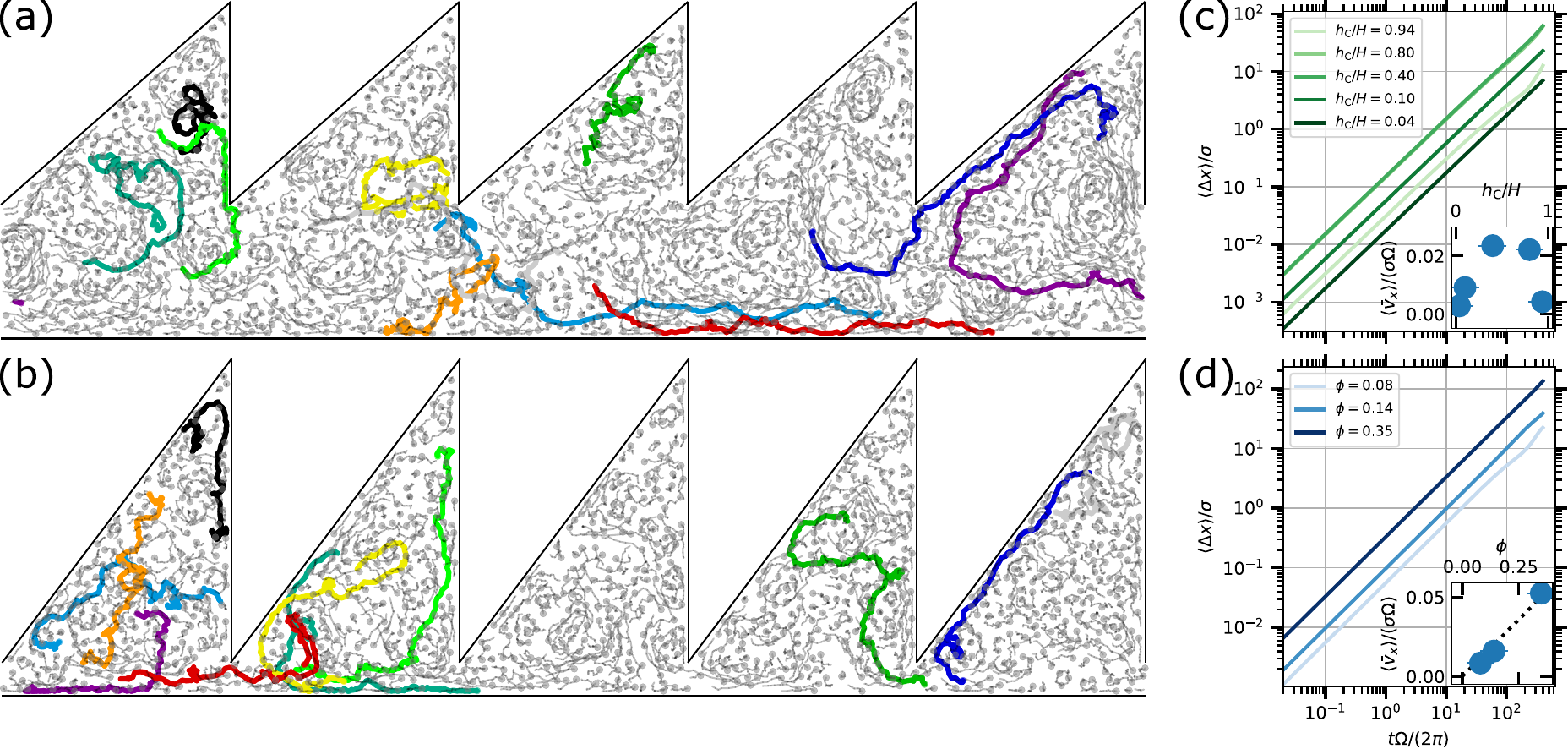}
	\caption{
		Vortex trapping and edge mode transport. %
		(a)--(b)~Simulation snapshots with rotor trajectories of duration $T\Omega/(2\pi) = 3$  with ten randomly selected rotor trajectories of duration $T\Omega/(2\pi) = 30$ highlighted in colour, for channels with channel-to-teeth separations (a)~$h_\mathrm{C}/H=0.4$ and (b)~$h_\mathrm{C}/H=0.1$, in systems with particle densities $\phi = 0.2$. %
		(c)--(d)~Time and ensemble averaged mean displacement along the channel for channels with (c)~various $h_c$, and (d)~various $\phi$.
	}
	\label{fig:fig2}
\end{figure}

The presence of the confinement interferes with the inherent dynamical vortex formation in two ways. %
On the one hand, steric interactions between the rotors and the walls locally restrict the vortex formation to the compartment between two teeth. %
On the other hand, the robust edge current along the walls leads to a stationary flow profile that superimposes to the inherent active turbulent behaviour. %
Simulation snapshots with superimposed rotor trajectories are shown in Fig.~\ref{fig:fig2}a--b. It can be seen how the rotors engage in the wall-restricted vortices and remain in the same compartment until they eventually escape before engaging in another spontaneously formed vortex. %
When the rotors get close enough to the walls, they get mostly engaged in the edge current featuring an almost straight trajectory along the wall. 
In this way, individual rotors close to the straight wall are guided into the direction determined by their chirality, here defined as positive $x$. %
When approaching the ratcheted wall, the rotors are effectively transported into the negative $x$-direction.
Since the edge velocity along each of the walls, $v_\mathrm{edge} $, is constant, the effective modulus of velocity along the channel is different in both walls.  
Rotors close to the ratcheted wall displace on average with $|v_{x}|\simeq  v_\mathrm{edge} \cos\theta$ which is by construction lower than the effective displacement in the positive $x-$direction for the rotors close to the flat wall, resulting in a net flux. Moreover, individual rotors at the ratcheted wall are more likely of getting temporarily trapped in tooth-compartment's vortices, enhancing the net flux. 

While individual rotors' $x$-displacement might show an oscillating behaviour, upon ensemble and time averaging, the mean $x$-displacement (see Fig.~\ref{fig:fig2}c--d) evolves ballistically into the positive $x$-direction $\langle x_i(t + t_0) - x_i(t_0) \rangle_{i,t_0} \sim \langle \bar{v}_x \rangle t$. %
For $h_\mathrm{C}/H \to 1$, we recover a linear flat (symmetric) channel, such that the individual rotors' average velocity $\langle \bar{v}_x \rangle \to 0$. %
On the other hand, for $h_\mathrm{C}/H \to 0$, the rotors get frequently trapped in spontaneous vortices within the same compartments and we also obtain $\langle \bar{v}_x \rangle \to 0$. %
For intermediate channel heights, we obtain the optimal tradeoff between the necessary wall asymmetry for net flux generation and hindrance introduced by the walls. We thus obtain a maximum average rotor velocity $\langle \bar{v}_x \rangle$ for $h_\mathrm{C} \simeq H/2$, as shown in Fig.~\ref{fig:fig2}c for averages computed over 24 individual simulations each of length $T\Omega/(2\pi) = 400$. %
The density of rotors in the system also plays a relevant role since the injected activity and the strength of the edge current increases linearly with rotor density, as described in  Eq.~\eqref{eq:v_edge}. Simulations also confirm that the average particle velocity $\langle \bar{v}_x \rangle$ is proportional to $\phi$, see Fig.~\ref{fig:fig2}d, being directly related to the emergent edge flow. %

\subsection{Emergence of material transport by robust edge modes}
\begin{figure}[h]
	\centering
	\includegraphics[width=.7\textwidth]{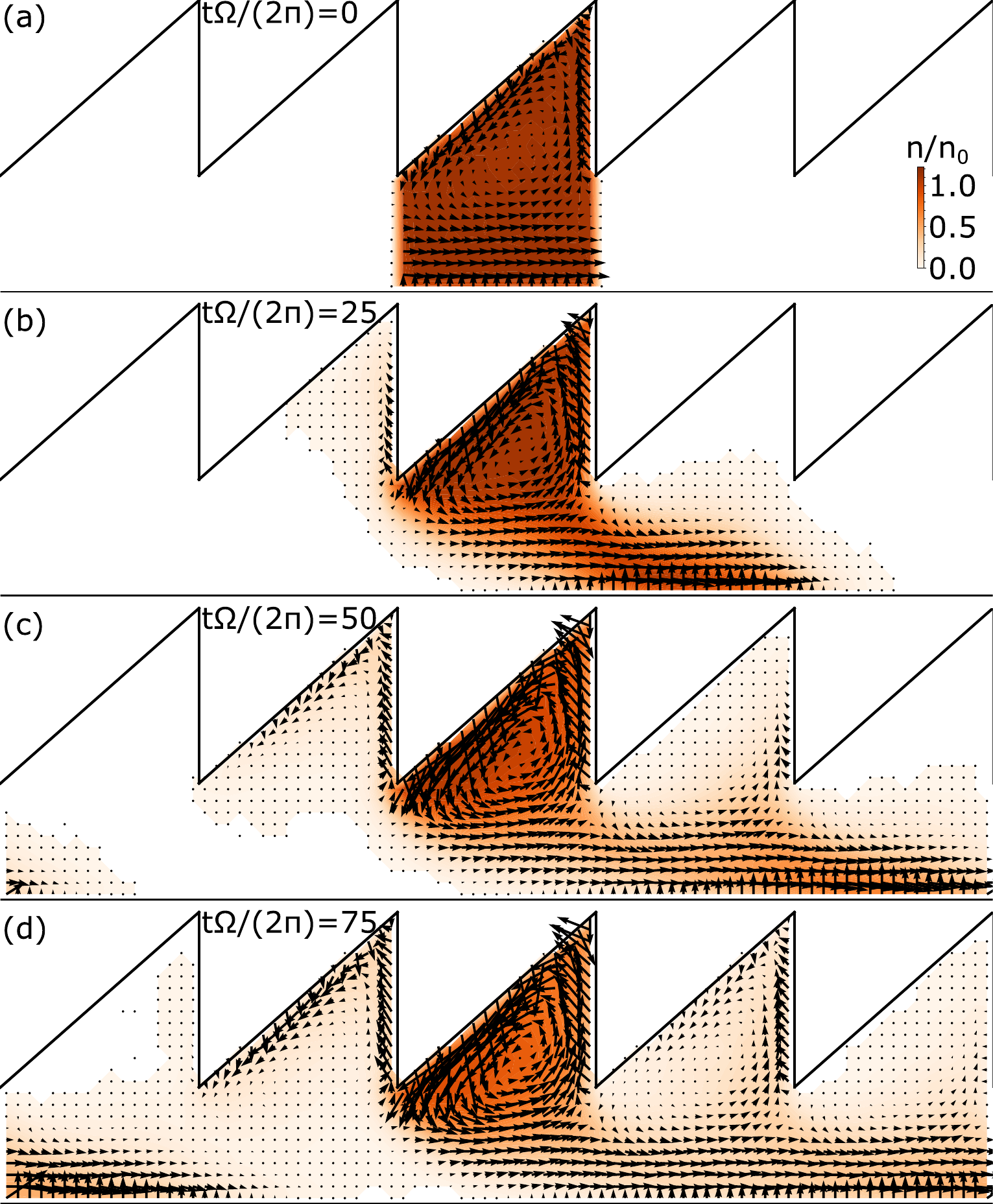}
	\caption{
		Edge mode facilitated net flux along the channel. %
		Evolution of tagged colloidal density in a system of constant density $\phi=0.35$. %
		The density field (heatmap) is obtained by counting the rotors per square bin of dimension $l_0 = 2\sigma$. %
		The diffusive flux density (arrows) is obtained by calculating the amount of rotors leaving the square bins into the neighbouring bins per unit time and unit bin-length. %
		The flux and density fields are time- and ensemble-averaged over 12 independent simulations each of length $T_\Omega/(2\pi) = 400$ with a time increment of $\Delta t/(2\pi) = 1.5$. %
	}
	\label{fig:fig5}
\end{figure}

The net flux along the channel is a result of the edge modes in chiral active systems. In order to observe the robustness of the edge transport phenomenon, we tag all rotors initially in one compartment and study their evolution with time. %
The density and flux of all tagged rotors in one compartment at time $t=0$ is shown in Fig.~\ref{fig:fig5}a, and  Figs.~\ref{fig:fig5}b--d display the diffusive spreading of the tagged rotor density at later times, \textit{i.e.}, how the tagged rotors spread in the channel. %
The transversal interactions among the rotors do not only lead to the emergence of odd viscosity~\cite{mecke2023simultaneous}, but also to odd diffusivity~\cite{hargus2021odd, mecke2025obstacle} leading to flux densities not only into the direction of the density gradient but also perpendicular to it. Particularly, a no-flux boundary condition leads to a transversal flux phenomenon along the wall even for vanishing density gradients~\cite{kalz2022collisions}. %
Note, that Fig.~\ref{fig:fig5} displays just the evolution of the tagged rotors, but the whole system has an approximately constant rotor density and thus constant even and odd diffusive contributions. %
We observe the emergence of flux densities along the walls resulting from no-flux boundary conditions. %
The resulting field $\bm{j}(\bm{r})$ bears a clear sign of the inherent transverse interactions and shows the emergence of a long-lived vortex structure in the centre of the compartment, driven by the edge flux and diffusive flux perpendicular to density gradients. %
The density around the vortex flux only slowly decreases as time progresses and the rotors predominantly leak out of the initial compartment in flux densities along the walls, see Fig.~\ref{fig:fig5}b. This is in accordance with the description of the rotors' trajectories, \textit{i.e.}, that rotors either tend to dwell in the vortices within a compartment or engage in the edge transfer into neighbouring compartments. %
The edge flux is almost scatter free accurately navigated around corners\textemdash a strong sign of topological protection~\cite{dasbiswas2018topological} of the edge mode (Fig.~\ref{fig:fig5}b--c). %
As time proceeds the (even) diffusive contributions lead to a spread out of the tagged rotors. Nevertheless, we observe that once a diffusively spread density element approaches another wall, it will again be reliably transported along the wall (Fig.~\ref{fig:fig5}c--d).
Although we observe rotor transport at both walls, the transport at the flat wall is considerably faster. While the rotors leaked out at the top of the channel travelled approximately one compartment to the left $\Delta x \approx -l_\mathrm{T}$, the rotors at the lower wall have travelled a distance of almost five compartments $\Delta x \approx 5l_\mathrm{T}$. This behaviour is a direct consequence of the net transport along the channel, which we complementarily quantify on the individual particle level $\langle x_i(t + t_0) - x_i(t_0) \rangle_{i,t_0}$, the emergent flow profile $\langle j_x \rangle$, as well as the odd diffusive density evolution. %

\subsection{Coarse-grained hydrodynamic flow fields and friction effect}
\begin{figure}[h]
	\centering
	\includegraphics[width=.75\textwidth]{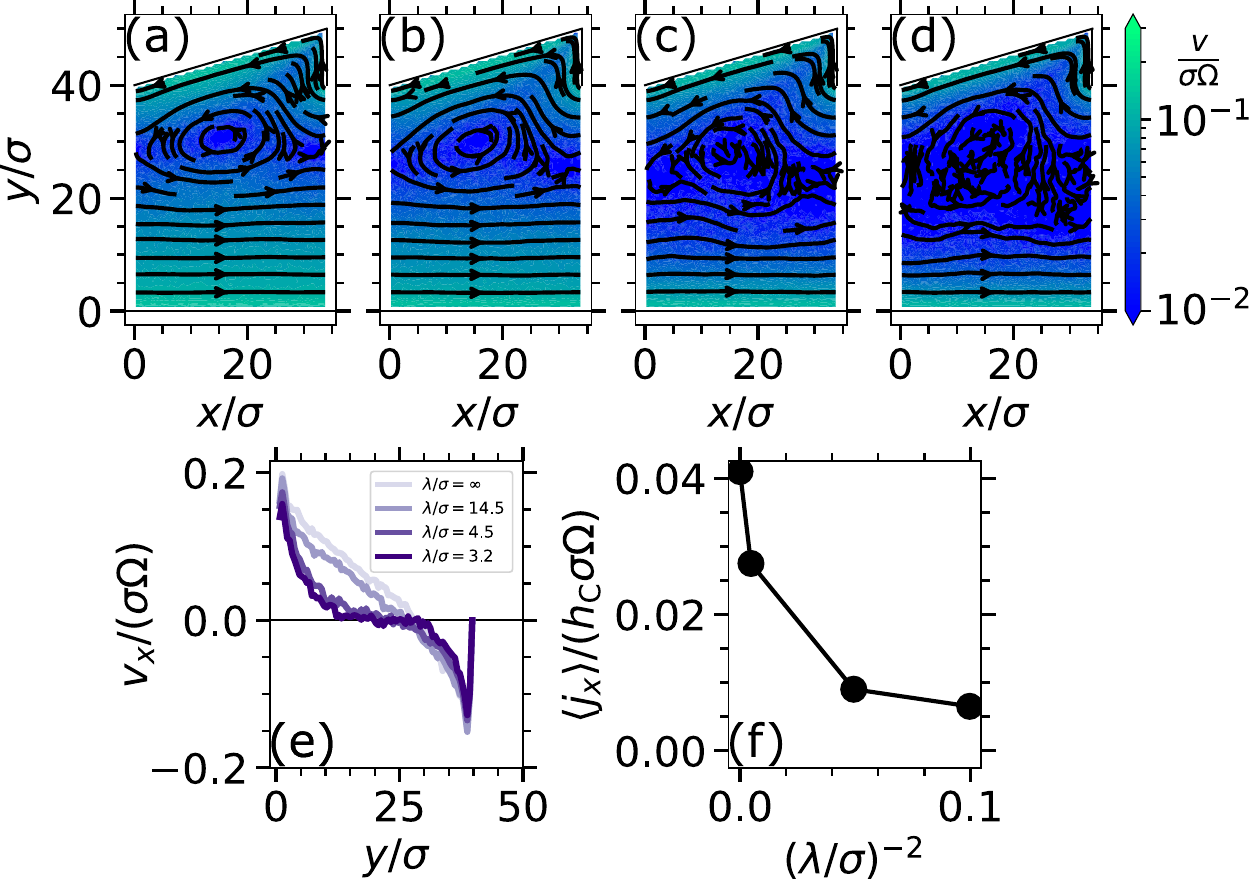}
	\caption{Average velocity flow fields, profiles and fluxes and friction effect.
		(a)--(d)~Magnitude of the averaged rotors velocity field $\bm{v}(\bm{r})$ (heatmap) with superimposed streamlines for varying values of the substrate friction and associated friction scale, (a)~$\lambda\to\infty$ (no friction), (b)~$\lambda=14.5\sigma$, (c)~$\lambda=4.5\sigma$, (d)~$\lambda=3.2\sigma$. %
		(e)~Corresponding velocity profiles measured at the compartment connecting position $v_x(x=0,y)$ for the different frictions. 
		(f)~Associated local net flux measured with Eq.~\eqref{eq:jx}. 
	}
	\label{fig:fig6}
\end{figure}

The spontaneously formed vortices discussed in Fig.~\ref{fig:fig2} have a very dynamic behaviour changing constantly size and position, such for understanding the persistent flow fields generated in the system we calculate the colloidal flow fields $\bm{v}(\bm{r})$ by time-averaging the rotor velocities in square grid defined by dividing the channel into bins of dimension $l_0=\sigma/2$. %
Figure~\ref{fig:fig6}a shows steady-state velocity flow fields with clear vortices, result of the edge flow, viscous coupling, and the continuity of the fields~\cite{mecke2024emergent}. We clearly can confirm the formation of a robust edge mode which propagates into the interior of the channel, as well as a persistent flow along the flat wall. %
Both flows decay away from the walls and result in a stagnation point, which, due to the ratcheted structure, does not lie in the centre of the channel. %

To more accurately mimic the flows in experimental setups, we perform simulations with various values of the friction coefficient $\gamma$ and associated friction decay length $\lambda = \sqrt{\nu/\gamma}$, as shown in 
Figs.~\ref{fig:fig6}a--d.  
The substrate friction leads to an increasing overall attenuation of the created flows, which increase the size of the vortices, specially of the internal stagnation areas. %
A quantification of the corresponding flow profiles into the $x$-direction are displayed in Fig.~\ref{fig:fig6}e, where it can be observed that the effect of the friction is stronger closer to the walls. %
While in the absence of substrate friction the flow profile bears only weak deviations from a linear profile resulting from the channel asymmetry, with decreasing $\lambda$ the flows decay faster. For the largest friction considered, $\lambda/\sigma = 3.2$, the edge flow decays to zero already at $y \approx 10\sigma$ and the stationary flow profile completely vanishes in the centre of the channel. %
Since the edge flow is a direct consequence of the stresses localised at the wall, the influence of the substrate friction on the strength of the edge flow directly at the boundary are minimal. %

The efficiency of the pumping capacity of these ratcheted channels can be quantified by characterising the local net flux along the channel, $j_x$. We integrate the flow into the $x$-direction over the channel height, 
\begin{align} \label{eq:jx}
	j_x = \int_0^{h_\mathrm{C}}\mathrm{d}y \, \hat{e}_x \cdot \bm{v} \, , 
\end{align}
which can be evaluated at one or various positions along the channel. 
Fig.~\ref{fig:fig6}f shows $\langle j_x \rangle$ relative to the channel height, as a function of the friction decay length, where a clear decrease with friction can be observed. %
Since the net flux is a result of the edge flow, and despite of the decrease with increasing friction it is important to note that it is nonzero for all considered values of $\lambda$ proving that the pumping effect is robust for a large range of experimental conditions where the substrate friction can often not be tuned easily. %

\subsection{Role of the channel degree of asymmetry}
\begin{figure}[h]
	\centering
	\includegraphics[width=\textwidth]{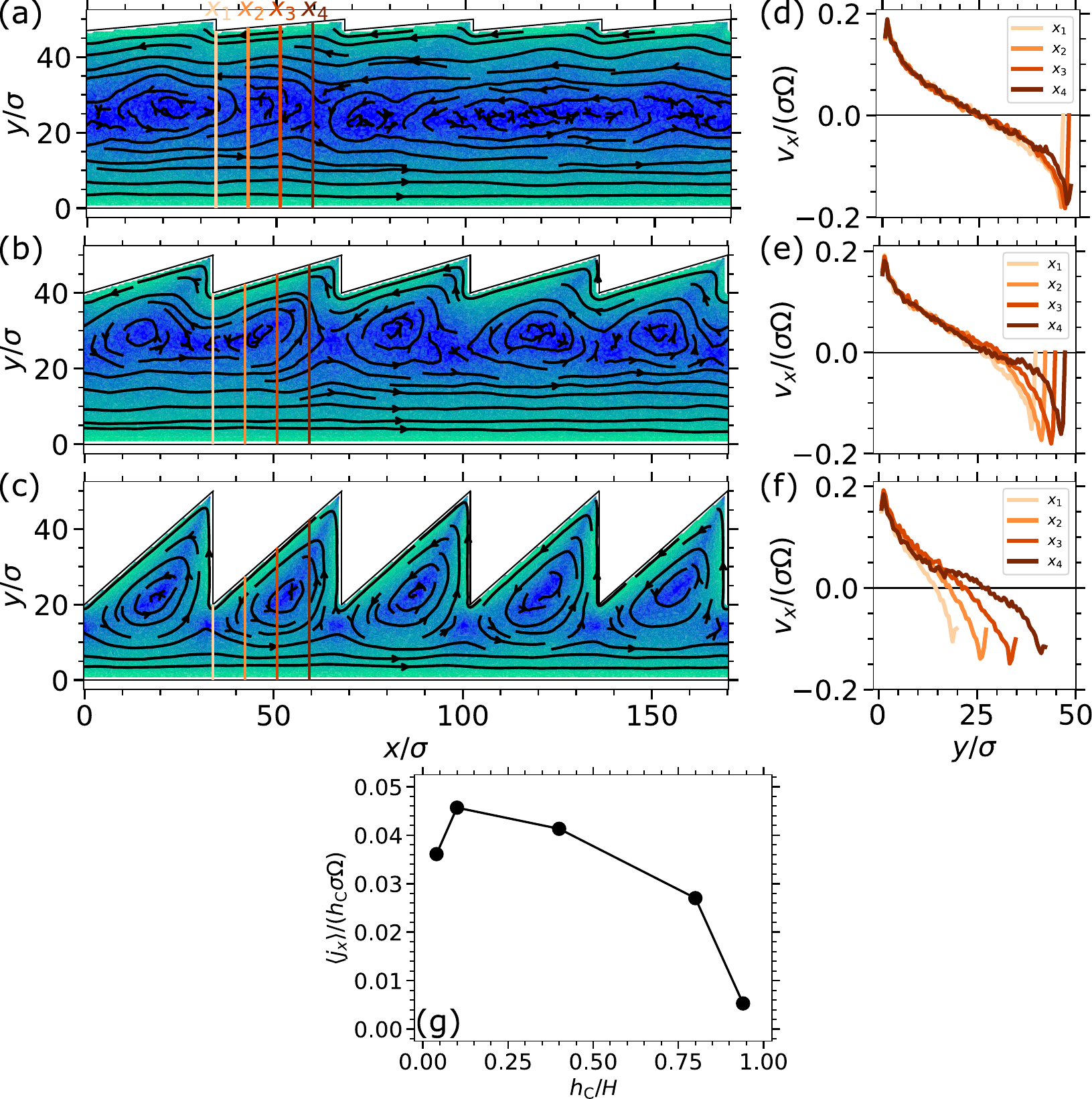}
	\caption{
		Average velocity flow fields, profiles and fluxes for channel with different symmetries. %
		(a)--(c)~Magnitude of the time-averaged edge flux generated emerging colloidal flow $\bm{v}(\bm{r})$ (heatmap) with superimposed streamlines for channels with various channel-to-teeth separations
		(a)~$h_\mathrm{C}/H=0.94$, (b)~$h_\mathrm{C}/H=0.80$, (c)~$h_\mathrm{C}/H=0.40$. %
		(d)--(f)~Corresponding velocity profiles $v_x(y)$ measured at different $x$-positions $x_1$, ..., $x_4$ as marked in (a)--(c). 
		(g)~Normalised mean flux density $\langle j_x \rangle$ measured with Eq.~\ref{eq:jx} 
		for various channel-to-teeth separations~$h_\mathrm{C}$. %
	}
	\label{fig:fig4}
\end{figure}

The asymmetry of the channel clearly determines the efficiency of the emergent pumped flow. %
Here, we keep the one wall asymmetrically ratcheted structure and modify the asymmetry by changing the channel-to-teeth separation $h_\mathrm{C}$, with fixed channel height $H$, what also modifies the teeth height $h_\mathrm{T}$, see Fig.~\ref{fig:fig1}c.  
Various emergent flows are shown in Fig.~\ref{fig:fig4}a--c, where in principle emergent vortices in each compartment can be observed. %
For $h_\mathrm{C} \to H$, \textit{i.e.}, for approaching a flat symmetric channel, this effect weakens, as shown in Fig.~\ref{fig:fig4}a. While the centre of the stationary vortex tends to be in the middle of the compartment into the $x$-direction $x_\mathrm{c} \approx (i+0.5)l_\mathrm{T}$, $i=0,1,...$, the $y$-position of the vortex centre happens to be displaced from the centre. %
Increasing the degree of asymmetry is here equivalent to decrease $h_\mathrm{C}$, as shown in Figs.~\ref{fig:fig4}b,c. This increase of asymmetry shows to increase the size of the vortices, as well as better pin their location inside the compartments. %
In all these cases, due to the different distances to the walls above and below the vortex, the amount of rotors transported above and below the vortex is uneven, giving rise to net transport along the channel. %
Moreover, the repetition of stationary vortices into the $x$-direction gives rise to a tank treading behaviour, where the active fluid layers between the stationary vortices and the lower (flat) wall are pumped into the $x$-direction, similar to a conveyor belt. %

We quantify the flow into the $x$-direction at various equally spaced positions in the compartments for varying $y$, see Fig.~\ref{fig:fig4}d--f. %
For $h_\mathrm{C} \to H$, the flow becomes point-symmetric about $y=H/2$ and the stagnation point is at $y \to H/2$. %
For decreasing $h_\mathrm{C}$, the stagnation point is shifted to the positive $y$-direction, see Figs.~\ref{fig:fig4}e--f. %
Figure~\ref{fig:fig4}g shows $\langle j_x \rangle$ per unit channel height as a function of $h_\mathrm{C}$. %
We observe a net flux into the $x$-direction for all studied $h_\mathrm{C}$. %
As the channel-to-teeth separation decreases, the stagnation point is shifted to larger $y$-values, and the net flux along the channel increases. However, when the channel-to-teeth separation becomes very small and comparable with the rotors size ($h_\mathrm{C} \to \sigma$), the gate from compartment to compartment becomes so narrow such that the steric repulsion between the rotors and the walls slows down and eventually prevents transport along the channel, such that we obtain maximum flux per unit channel height for $h_\mathrm{C} = 5\sigma$. %

\subsection{Role of channel-chirality inversion}

\begin{figure}[h]
	\centering
	\includegraphics[width=.9\textwidth]{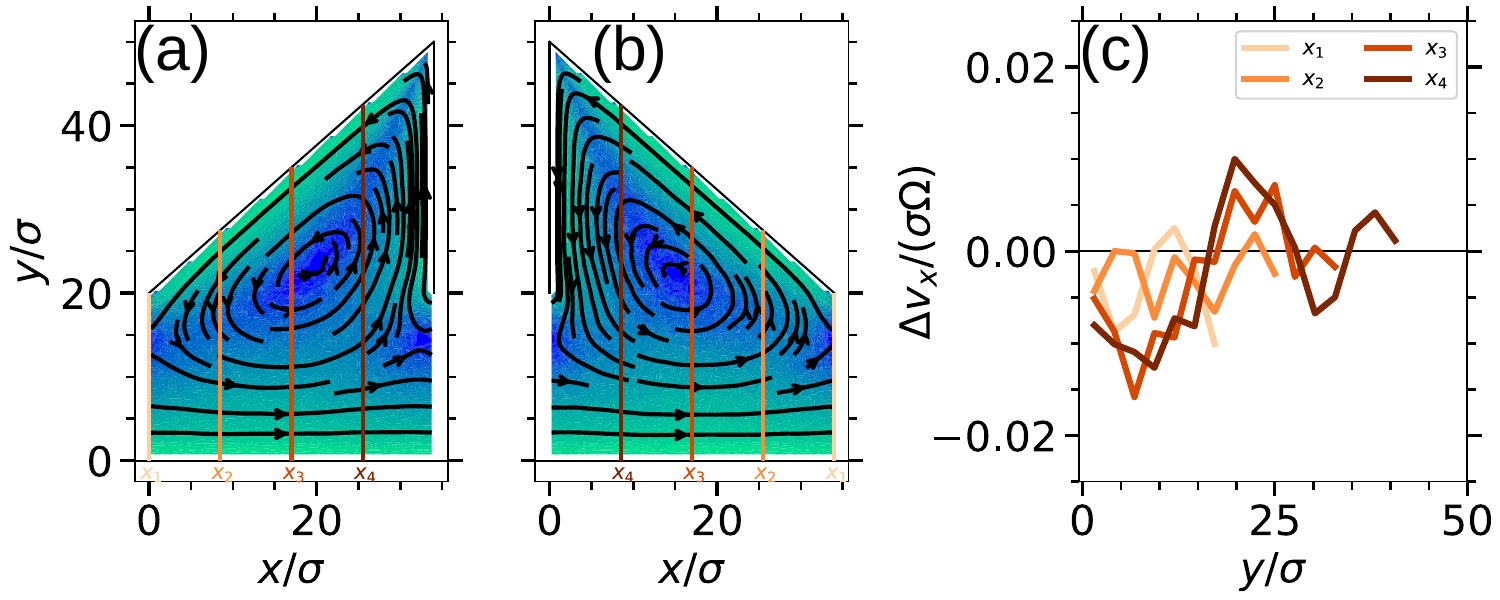}
	\caption{
		Differences in the emergent flow of ratcheted channels with opposite inclinations, here denoted as R in (a) and L in (b). %
		(a,b)~Magnitude of the average flow with superimposed streamlines for a system corresponding to R and L, respectively. Colour code as in Fig.~\ref{fig:fig6}. %
		(c)~Difference in flow profiles $\Delta v_x(x_i,y) = v_x^{\mathrm{R}}(x_i,y) - v_x^{\mathrm{L}}(x_i,y)$ at the marked positions in (a) and (b).
	}
	\label{fig:cinv}
\end{figure}

Given the symmetry of the system, it is clear that reversing both the rotation direction of the rotors and the direction of the the asymmetric ratcheted wall will reverse the direction of the flow and make the system exactly mirror symmetric. 
What \emph{a priori} is not so evident is what happens if only one of the two is reversed.
To investigate this aspect we perform simulations where the inclination of the ratchet is reverted, we refer to the two geometries as R, L  as shown in Fig.~\ref{fig:cinv}a,b.
The direction of the edge mode solely depends on the chirality of the active fluid, such that in Fig.~\ref{fig:cinv}a the flows are moving upwards along the vertical wall segment into the acute angle, while in Fig.~\ref{fig:cinv}b, the created flows are coming out of the acute angle along the wall. %
To better quantify this, Fig.~\ref{fig:cinv}c shows the difference of the emergent flows at marked positions $x_{1,2,3,4}$ of equal local channel height $\Delta v_x(x_i,y) = v_x^{\mathrm{R}}(x_i,y) - v_x^{\mathrm{L}}(x_i,y)$, where $v_x^{\mathrm{R}}$ and $v_x^{\mathrm{L}}$ correspond to Fig.~\ref{fig:cinv}a and b, respectively. %
We observe subtle changes in the flow into the $x$-direction, particularly when approaching $x$-positions close to the acute wall angle. %
On the individual rotor level, we calculate the mean $x$-displacement and extract the mean velocities. We obtain $\langle \bar{v}_x^{\mathrm{R}} \rangle = 0.024$ and $\langle \bar{v}_x^{\mathrm{L}} \rangle = 0.027$, which constitutes a difference of about 10\%. %
This means that quantitatively both geometries are not identical, but very similar. The geometry where the flow created at the long segment of the ratcheted wall ends against the short segment, see Fig.~\ref{fig:cinv}b, leads to slightly higher net fluxes than the the geometry where the flow created at the long segment of the ratcheted wall ends in the centre of the channel, see Fig.~\ref{fig:cinv}a. 

\section{Discussion}

Chiral active systems in bulk are known to generate dynamic circulating flows and related active turbulence. Here we have shown how asymmetric confinement can solely convert the individual particle microscopic rotational motion into a directed macroscopic transport. 
When a dense system of rotors get close to a wall, unidirectional rotational stresses lead to the formation of a robust edge current and subsequently to a stationary flow profile which superimposes to the inherent active turbulent behaviour of the rotors.  
In the case of asymmetric confinement, the effective velocity of the rotors along the channel walls has not only opposite direction, but it is also asymmetric in modulus, what generates a net flux. Furthermore, the asymmetry between the confining walls leads to stagnation points of the created flows displaced from the centre of the channel admitting net flux densities along the channel. %
\begin{figure}[h]
	\centering
	\includegraphics[width=\textwidth]{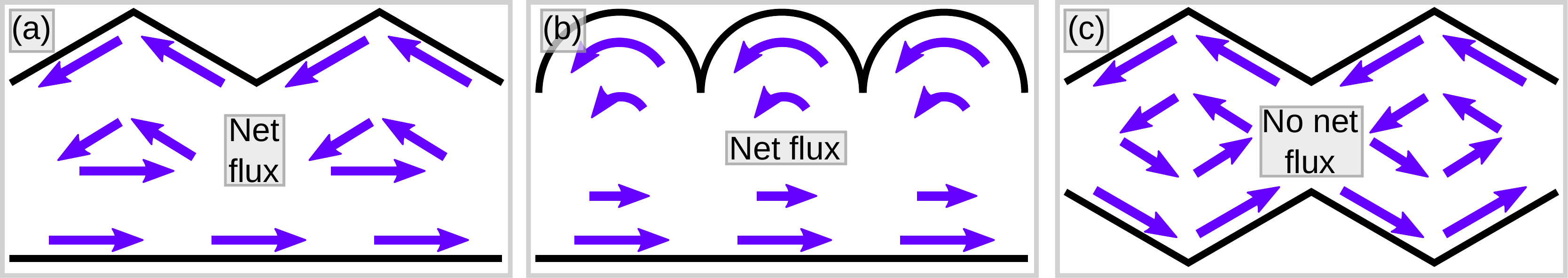}
	\caption{
		Sketches of three differently patterned channels together with the associated expected flow fields and fluxes. 
		a)~Flat wall with opposite symmetric ratcheted wall, 
		b)~flat wall with opposite periodically curved  wall,
		c)~double symmetric ratcheted wall. 
	}
	\label{fig:fig7}
\end{figure}
In this work, we have investigated in detail a channel geometry consisting of a flat wall and an opposite asymmetric ratcheted. The asymmetry of the ratchet is though not necessary in order to generate net transport. Other flux generating geometries are for example  channels consisting out of a flat wall and a symmetrically ratcheted wall, {\em i.e.} with teeth of equal inclination on both sides of the ratchet teeth, as illustrated in Fig.~\ref{fig:fig7}a, or a geometry of consecutive curved compartments in only one wall, as illustrated in Fig.~\ref{fig:fig7}b. These geometries are also expected to generate vortices pinned to each of the tooth-compartments and stagnation points displaced from the centre of the channel which will favor the flow velocity along the flat wall. 
The more symmetric structures might pin the vortices a bit less effectively than the asymmetric ratcheted ones, which could result in sightly less effective fluxes, although this extent requires further investigation. Another difference of the more symmetric structures in Fig.~\ref{fig:fig7}a,b is that if the colloids reverse the direction of rotation, the flows and related flux will simple reverse in direction making the pumping exactly reversible, which can be important in practical applications. %
Similarly if a channel has both walls non-flat, but similarly patterned, the system is not expected to result in an effective flux, as show in the example in Fig.~\ref{fig:fig7}c for a double symmetric ratcheted channel. %

In conclusion, using mesoscale hydrodynamic simulations, we demonstrate that intrinsically asymmetric channels give rise to the spontaneous formation of a persistent net flow in the absence of externally applied pressure gradients or body forces. Our findings are expected to extend to a broad range of geometrically confined systems, providing a versatile and efficient mechanism for converting chiral activity into autonomous fluid transport. This approach offers a promising route toward the development of self-powered microfluidic pumps.

%
%


\ack{J.M. acknowledges the National Natural Science Foundation of China for supporting this work within the Research Fund for International Young Scientists under Grant No. 12350410368. %
J.M. and M.R. gratefully acknowledge the Gauss Centre for Supercomputing e.V. (www.gauss-centre.eu) for funding this project by providing computing time through the John von Neumann Institute for Computing (NIC) on the GCS Supercomputer JUWELS at J\"ulich Supercomputing Centre (JSC) and the Helmholtz Data Federation (HDF) for funding this work by providing services and computing time on the HDF Cloud cluster at the J\"ulich Supercomputing Centre (JSC).
Y.G. acknowledges funding support from the Natural Science Foundation of Guangdong Province (2024A1515011343) and from the Key Project of Guangdong Provincial Department of Education (2023ZDZX3021).}


\data{The data that support the findings of this study are available from the corresponding author upon reasonable request.}


\bibliographystyle{unsrt}
\bibliography{Bibliography}

\end{document}